\documentclass[12pt]{article}

\usepackage{amsmath,amssymb,bm,graphicx}
\usepackage{color} 
\usepackage{hyperref}

\newcommand{\pslash}{\not \! p}

\newcommand{\Bslash}{\not \! B}

\begin{document}

\vskip 0.5 truecm

\begin{center}
{\Large{\bf Neutron-antineutron oscillation\\\vskip 0.5 cm and parity and CP symmetries}}
\end{center}
\vskip .5 truecm
\begin{center}
{\bf { Kazuo Fujikawa$^\dagger$ and Anca Tureanu$^*$}}
\end{center}

\begin{center}
\vspace*{0.4cm} 
{\it {$^\dagger$ Quantum Hadron Physics Laboratory, RIKEN Nishina Center,\\
Wako 351-0198, Japan\\
$^*$Department of Physics, University of
Helsinki, P.O.Box 64, \\FIN-00014 Helsinki,
Finland\\
}}
\end{center}

%\makeatletter
%\@addtoreset{equation}{section}
%\def\theequation{\thesection.\arabic{equation}}
%\makeatother

\begin{abstract}
   It is  shown that the neutron-antineutron oscillation {\em per se} does not necessarily imply CP violation in an effective local Lorentz invariant description of the neutron which preserves CPT, contrary to a recent analysis in the literature. 
%This would indeed be very interesting and an addition  to
%the Kobayashi-Maskawa phase for CP violation in such a  $\Delta B = 2$ interaction. 
A CP- and baryon number-violating term can be transformed into a CP conserving one, thus rendering the CP violation spurious and unusable in the analysis of baryogenesis, for example. 
It is also shown that the neutron-antineutron oscillation in
a $\Delta B=2$, Lorentz and CPT invariant  interaction  can occur only when  parity is violated, irrespective of the CP properties; when parity is conserved, it is the ordinary quantum transition from neutron to antineutron which takes place.
Those statements are proven by explicitly analyzing all the possible
combinations  with P, C and CP violation or their conservation.
Moreover, a suitable combination of P=odd  and P=even $\Delta B=2$ interactions in the present model is shown to give rise to the CP preserving mass term for the right-handed neutrino if one replaces the neutron by the neutrino, reinforcing the conclusion. 
\end{abstract}
%\maketitle
%\large
 
\section{Introduction}

It has been argued recently that the observation of the neutron oscillation inevitably implies  violation of CP symmetry~\cite{berezhiani}. If this statement is confirmed 
it would imply many interesting physical consequences. We here examine this issue in more detail and show that the neutron-antineutron oscillation {\em per se} does not necessarily imply CP violation. This aspect is closely related to the CP preserving mass term for the 
right-handed neutrino.
It is also shown that the parity symmetry plays a crucial role in the occurrence of neutron-antineutron oscillation in a conventional sense in Lorentz invariant local effective theory.  

Following Ref.~\cite{berezhiani}, we start with the  free neutron Lagrangian  defined by 
\begin{eqnarray}\label{1}
{\cal L}&=&\overline{n}(x)i\gamma^{\mu}\partial_{\mu}n(x) - m\overline{n}(x)n(x),
\end{eqnarray}
that is invariant under the global phase rotation
\begin{eqnarray}\label{2}
n(x)\rightarrow e^{i\alpha}n(x), \ \ \ \ \overline{n}(x)\rightarrow e^{-i\alpha}\overline{n}(x),
\end{eqnarray}
which defines the notion of {\em  baryon number}. See also Refs.~\cite{marshak, chang, krivoruchenko} for closely related earlier works.

The baryon number violating term with $\Delta B=2$, which is {\em hermitian}, 
 is defined in Ref.~\cite{berezhiani} by 
\begin{eqnarray}\label{3}
{\cal L}_{\Bslash}=-\frac{1}{2}\epsilon[n^{T}(x)Cn(x)+ \overline{n}(x)C\overline{n}^{T}(x)]
\end{eqnarray}
where $\epsilon$ is a {\em real} number and C is the charge conjugation matrix. One can add a parity violating mass term to the above Lagrangian in \eqref{1},
\begin{eqnarray}\label{4}
{\cal L}_{PV}= im^{\prime}\bar{n}(x)\gamma_{5}n(x),
\end{eqnarray}
which is invariant under the above phase transformation and thus baryon number conserving, but it violates P and CP, as well as the time reversal symmetry T. It has been noted that this parity violating mass term is not eliminated by a global chiral transformation
if one wishes to keep the baryon number violating term in \eqref{3} intact~\cite{berezhiani}.
The set of equations \eqref{1} to \eqref{4} are the basis of the analysis in~\cite{berezhiani}. We also adopt the effective Lorentz invariant local description of the neutron in the present paper. See references in~\cite{mohapatra} for the original proposals of the neutron oscillation, which is reviewed in~\cite{reviews}. The present experimental status is found in~\cite{experiment}.

One can confirm that the starting Lagrangian \eqref{1} is invariant under all the discrete transformations P, C, and T (and thus CPT also), while the above baryon number violating term $\int d^{4}x{\cal L}_{\Bslash}$ in \eqref{3} satisfies 
\begin{eqnarray}\label{5}
P=odd, \ \ \ \  C=even, \ \ \ \  T=odd, \ \ \ \  CPT=even,
\end{eqnarray}
and thus 
\begin{eqnarray}\label{6}
CP=odd,
\end{eqnarray}
which was emphasized in~\cite{berezhiani}. 
In the Appendix, we briefly summarize the definitions of the various discrete transformations we use.

We now observe that, if one performs a $\pi/4$ phase rotation
\begin{eqnarray}\label{7}
n(x)\rightarrow e^{i\pi/4}n(x),
\end{eqnarray}
one obtains 
a {\em hermitian} baryon number violating term from ${\cal L}_{\Bslash}$ in \eqref{3}, 
\begin{eqnarray}\label{8}
{\cal L}^{\prime}_{\Bslash}=-\frac{i}{2}\epsilon[n^{T}(x)Cn(x) - \overline{n}(x)C\overline{n}^{T}(x)].
\end{eqnarray}
Remark that the baryon number preserving Lagrangians in \eqref{1} and \eqref{4} are invariant under this phase transformation, and one may naively expect that once we choose a very small baryon number violating term in \eqref{3}, the C and CP symmetry properties would not change under a phase transformation.
However,
one can confirm that $\int d^{4}x{\cal L}^{\prime}_{\Bslash}$ satisfies 
\begin{eqnarray}\label{9}
P=odd, \ \ \ \  C=odd, \ \ \ \  T=even, \ \ \ \  CPT=even,
\end{eqnarray}
and thus
\begin{eqnarray}\label{10}
CP=even.
\end{eqnarray}
This shows that the charge conjugation property of the  baryon number violating term has no definite meaning when one performs the phase transformation, which defines the baryon number. The baryon number violating term implies that what is the particle and what is the anti-particle is not unequivocal. The CP property of the baryon number violating term is {\em ill-defined}, and thus it may not be used in the physical analysis of baryogenesis, for example. Technically,  this uncommon feature arises from the fact that C is specified for {\em a combination of two terms} in \eqref{3}, while P and T are specified for each term in \eqref{3} separately. One may conclude that the observation of the neutron oscillation {\em per se} does not necessarily imply CP violation, contrary to the analysis in~\cite{berezhiani}. 

As is noted in~\cite{berezhiani}, one cannot eliminate the CP violating term \eqref{4}
if such a term appears in the Lagrangian, without modifying \eqref{3}. However, if one takes the CP violation in \eqref{4} seriously, the Lagrangian \eqref{8} rather than \eqref{3} satisfies one of the three conditions of Sakharov for baryogenesis, which requires {\em both} C and CP violation.

\section{Various forms of $\Delta B=2$ interaction}

In some of the analyses in particle physics, it is common to fix the chiral transformation freedom by first eliminating the parity violating mass term \eqref{4} and then discuss other symmetries 
such as the possible neutron number violating interactions.
If one adopts this approach, one has the free Lagrangian in \eqref{1} and the baryon number violating interactions such as \eqref{3} and \eqref{8}. In addition to the interactions \eqref{3} and \eqref{8}, which are parity odd, one can also consider the parity-even neutron-number violating hermitian operators,
\begin{eqnarray}\label{11}
{\cal L}_{\gamma_{5}\Bslash}&=&-\frac{1}{2}\epsilon[n^{T}(x)C\gamma_{5}n(x) - \overline{n}(x)C\gamma_{5}\overline{n}^{T}(x)]\nonumber\\
&=&-\frac{1}{2}\epsilon[\overline{n^{c}}(x)\gamma_{5}n(x) - \overline{n}(x)\gamma_{5}n^{c}(x)]
\end{eqnarray}
which has
\begin{eqnarray}\label{12}
P=even, \ \ \  C=odd, \ \ \  CP=odd, \ \ \  T=odd,  \ \ \  CPT=even,
\end{eqnarray}
and thus satisfies one of Sakharov's conditions, which requires the breaking of both C and CP, as well as
\begin{eqnarray}\label{13}
{\cal L}'_{\gamma_{5}\Bslash}&=&-\frac{i}{2}\epsilon[n^{T}(x)C\gamma_{5}n(x) + \overline{n}(x)C\gamma_{5}\overline{n}^{T}(x)]\nonumber\\
&=&-\frac{i}{2}\epsilon[\overline{n^{c}}(x)\gamma_{5}n(x)+\overline{n}(x)\gamma_{5}n^{c}(x)]
\end{eqnarray}
which has
\begin{eqnarray}\label{14}
P=even, \ \ \  C=even, \ \ \  CP=even, \ \ \  T=even,  \ \ \  CPT=even, 
\end{eqnarray}
and thus preserves all the discrete symmetries.

The two interactions \eqref{11} and \eqref{13} are related to each other by a $\pi/4$ phase rotation of the neutron field, and the charge conjugation property of the baryon number violating term is again changed by the phase rotation.

In the next section we show explicitly that the neutron oscillation in the conventional sense does not take place in the parity preserving $\Delta B=2$ interactions.
What happens in the parity conserving case is that the oscillation time becomes infinite,
namely, there is no observable oscillation, but instead the neutron and antineutron states have a non-vanishing overlap in the presence of the $\Delta B=2$ interaction, and thus the quantum mechanical transition of the neutron to the antineutron can generally take place. However, signals characteristic to {\em oscillation} are absent.  This prominent role of parity in the analysis of neutron oscillation phenomena is interesting. 

\section{Explicit solutions}

The neutron oscillation is a subtle phenomenon and thus it is useful to solve explicitly our quadratic Lagrangian, which is regarded as describing asymptotic fields. Parity has a major role in distinguishing between oscillation and the lack of it, therefore we shall split the analysis with respect to the parity transformation properties of the baryon number violating term. Recall that the charge conjugation properties of the baryon number violating terms vary under a phase transformation, consequently the CP properties will vary as well. However, the total Lagrangian is local as well as Lorentz invariant, and thus CPT preserving in every case.

\subsection{P=odd,\ CP=even or odd}

We analyze the hermitian Lorentz invariant local Lagrangian consisting of the sum of \eqref{1} and \eqref{8},
but we allow for a phase for the baryon number violating parameter
\begin{eqnarray}\label{21}
\epsilon \rightarrow \epsilon e^{-i\alpha},\ \ \ \  \epsilon>0,
\end{eqnarray}
such that the total Lagrangian is written as
\begin{eqnarray}\label{20}
{\cal L}&=&\overline{n}(x)i\gamma^{\mu}\partial_{\mu}n(x) - m\overline{n}(x)n(x)
-\frac{i}{2}\epsilon[e^{-i\alpha}n^{T}(x)Cn(x) - e^{i\alpha}\overline{n}(x)C\overline{n}^{T}(x)].\nonumber\\
\end{eqnarray} 
Note that: \\
$i)$ case $\alpha=0$ corresponds to the P=odd, C=odd, CP=even Lagrangian in \eqref{8};\\
$ii)$  case $\alpha=\pi/2$ corresponds to the P=odd, C=even, CP=odd Lagrangian in \eqref{3}. 

It is instructive to consider a hermitian interaction Lagrangian
\begin{eqnarray}\label{22}
{\cal L}_{int}&=&-\frac{i}{2}[n^{T}(x)Cn(x)\phi(x) - \overline{n}(x)C\overline{n}^{T}(x)\phi(x)^{\dagger}],
\end{eqnarray}
where the transformation property of $\phi(x)$ is 
$\phi(x)\rightarrow e^{-2i\alpha}\phi(x)$ for $n(x)\rightarrow e^{i\alpha}n(x)$, P=odd (${\cal P}\phi(x){\cal P}^{-1}=-\phi(x)$), T=even (${\cal T}\phi(x){\cal T}^{-1}=\phi(x)$) and ${\cal CP}\phi(x)({\cal CP})^{-1}=\phi^{\dagger}(x)$ under CP. Thus, ${\cal L}_{int}$ is baryon number preserving and invariant under the discrete symmetries P, CP, and T. One may now consider the spontaneous symmetry breaking 
\begin{eqnarray}\label{23}
\langle 0,\alpha|\phi(x)|0,\alpha\rangle=\epsilon e^{-i\alpha},
\end{eqnarray}
to obtain the Lagrangian \eqref{20}. The vacuum $|0,\alpha\rangle$ parameterized by $\alpha$ breaks the baryon number symmetry together with P.
Under parity, $\epsilon\rightarrow -\epsilon$, and thus we would need in principle to consider the tensor product of Hilbert spaces, namely, $${\cal H}(\alpha)={\cal H}_{1}(\alpha,\epsilon)\otimes {\cal H}_{2}(\alpha,-\epsilon),\ \ \ \rm{with}\ \epsilon>0,$$ in order to realize the operator algebra consistently. We emphasize that physical proceses are defined in ${\cal H}_{1}(\alpha,\epsilon)$.

Note that C invariance requires 
\begin{equation}\label{23'}
(\epsilon e^{-i\alpha})^{\star}=-\epsilon e^{-i\alpha}.
\end{equation} 

We distinguish again the following cases:\\
$i)$ for $\alpha=0$, CP and T (and thus CPT) are preserved, while C is violated as condition \eqref{23'} is not fulfilled; \\
%%%%%%%%%%%
$ii)$ for $\alpha= \pi/2$, which corresponds to \eqref{3}, C is preserved, while CP and T are broken;\\
$iii)$ for $\alpha\neq 0$ and $\pi/2$, C, CP, and T are broken but CPT is preserved.

In view of the above discussion, we shall define the discrete transformations case by case, depending on which symmetries are preserved. Customarily, the antiparticle is defined by C-conjugation, but this definition makes sense only when the theory is C-invariant. In our framework, this is possible only for the case when $\alpha= \pi/2$. When C is violated, one has the option of defining the antiparticle by CP-conjugation, if CP is preserved, or ultimately by CPT-conjugation, if both C and CP are violated. Actually, the definition of the antiparticle by CPT transformation holds always, irrespective of the C and CP properties of the theory.

\paragraph {C-conserving case ($\alpha= \pi/2$):}

We obtain the equations of motion from \eqref{20}:
\begin{eqnarray}\label{24}
&&[i\gamma^{\mu}\partial_{\mu}-m]n(x)-\epsilon n^{c}(x)=0,\nonumber\\
&&[i\gamma^{\mu}\partial_{\mu}-m]n^{c}(x)-\epsilon n(x)=0,
\end{eqnarray}
with $n^{c}=C\overline{n}^{T}$, which are rewritten as 
\begin{eqnarray}\label{25}
[i\gamma^{\mu}\partial_{\mu}-m](n(x)\pm n^{c}(x))\mp \epsilon (n(x)\pm n^{c}(x))=0.
\end{eqnarray}
We define the combinations
\begin{eqnarray}\label{26}
\psi_{\pm}(x)=\frac{1}{2}[n(x)\pm n^{c}(x)],
\end{eqnarray}
which satisfy Dirac equations with {\em different masses}, 
\begin{eqnarray}\label{27}
[i\gamma^{\mu}\partial_{\mu}-(m\pm \epsilon)]\psi_{\pm}(x)=0.
\end{eqnarray}
We thus have
\begin{eqnarray}\label{29}
&&n(x)=\psi_{+}(x)+\psi_{-}(x),\nonumber\\
&&n^{c}(x)=\psi_{+}(x)-\psi_{-}(x),
\end{eqnarray}
inside  the vacuum $|0,\pi/2\rangle$. 
By comparing the definition of $n^{c}(x)=\psi^{c}_{+}(x)+\psi^{c}_{-}(x)$ with the second expression in \eqref{29} we obtain
\begin{eqnarray}\label{28}
&&\psi^{c}_{\pm}(x)=\pm\psi_{\pm}(x),
\end{eqnarray}
showing that $\psi_{\pm}(x)$ are Majorana fields. 

We  define the conventional classical solutions
\begin{eqnarray}\label{31}
&&[\pslash-(m\pm\epsilon)]u(\vec p,m\pm\epsilon,s)=0,\nonumber\\
&&[\pslash+(m\pm\epsilon)]v(\vec p,m\pm\epsilon,s)=0
\end{eqnarray}
%and note the relations (see Appendix)
%\begin{eqnarray}\label{32}
%u^{c}(\vec p,m\pm\epsilon,s)=v(\vec p,m\pm\epsilon,s), \ \  v^{c}(\vec p,m\pm\epsilon,s)=u(\vec p,m\pm\epsilon,s).
%\end{eqnarray}
and expand the Majorana fields as 
\begin{eqnarray}\label{33}
\psi_{\pm}(x)&=&\int \frac{d^{3}p}{(2\pi)^{3/2}}\sum_{s}\{a(\vec{p},\pm,s)u(\vec{p},\pm,s)e^{-ip_\pm x}+b^{\dagger}(\vec{p},\pm,s)v(\vec{p},\pm,s)e^{ip_\pm x}\},\nonumber\\
\end{eqnarray}
where we used the notation $u(\vec{p},\pm,s)\equiv u(\vec{p},m\pm\epsilon,s)$ and $p_\pm x=\sqrt{\vec{p}^{2}+(m\pm\epsilon)^{2}}x^0+\vec{p}\cdot\vec{x}$. This expansion is actually valid for the more general classes of fields appearing in \eqref{27'} and \eqref{33'}.

We impose the conventional anti-commutation relation $$\{n(x^{0},\vec{x}),n^{\dagger}(x^{0},\vec{y})\}=\delta^{3}(\vec{x}-\vec{y})$$ implied by the Lagrangian \eqref{20}, which 
is satisfied by 
\begin{eqnarray}\label{34}
&&\{a(\vec{p},\pm,s),a^{\dagger}(\vec{p^{\prime}},\pm,s^{\prime})\}=\frac{1}{2}\delta_{s,s^{\prime}}\delta^{3}(\vec{p}-\vec{p^{\prime}}),\nonumber\\
&&\{b(\vec{p},\pm,s),b^{\dagger}(\vec{p^{\prime}},\pm,s^{\prime})\}=\frac{1}{2}\delta_{s,s^{\prime}}\delta^{3}(\vec{p}-\vec{p^{\prime}}),\nonumber\\
&&\{a(\vec{p},+,s),a^{\dagger}(\vec{p^{\prime}},-,s^{\prime})\}=0, \ \ \ 
\{b(\vec{p},+,s),b^{\dagger}(\vec{p^{\prime}},-,s^{\prime})\}=0.
\end{eqnarray}
The relations \eqref{28} constrain the creation and annihilation operators, but those constraints are immaterial for our further applications.

Note that by C-conjugation the states of ${\cal H}_{1}(\alpha,\epsilon)$ are transformed among themselves. The same is valid for the CPT transformation. However, the P and CP transformations, as well as T, take states of ${\cal H}_{1}(\alpha,\epsilon)$ to states from ${\cal H}_{2}(\alpha,-\epsilon)$. For the analysis of oscillations, precise definitions of these transformation properties are irrelevant.

\paragraph{C-violating, CP-conserving case ($\alpha=0$):} In this situation, the P and C transformations interchange the vacuum states of the Hilbert spaces ${\cal H}_{1}(\alpha,\epsilon)$ and ${\cal H}_{2}(\alpha,-\epsilon)$, therefore they mix the states among these spaces. If we were to define the antiparticle in the mass eigenstates by C-conjugation, that antiparticle would be in the non-physical space. Consequently, we shall re-write the equations of motion only in terms of $n(x)$ and $n^{cp}(x)$ and assume only CP-transformation properties for the creation and annihilation operators, so that not to leave the physical Hilbert space.

Within the vacuum $|0,0\rangle$, the equations of motion read:
\begin{eqnarray}\label{24'}
&&[i\gamma^{\mu}\partial_{\mu}-m]n(t,\vec x)-i\epsilon \gamma^0n^{cp}(t,-\vec x)=0,\nonumber\\
&&[i\gamma^{\mu}\partial_{\mu}-m]\gamma^0n^{cp}(t,-\vec x)-\epsilon n(t,\vec x)=0,
\end{eqnarray}
with $n^{cp}(t,-\vec x)=-\gamma^0C\overline{n}^{T}(t,\vec x)$ (to be precise,
$({\cal CP})n_{\alpha}(t,-\vec x)({\cal CP})^{-1}=C_{\alpha\beta}n^{\dagger}_{\beta}(t,\vec x)$). We re-write \eqref{24'} as
\begin{eqnarray}\label{25'}
[i\gamma^{\mu}\partial_{\mu}-m](n(t,\vec x)\pm i\gamma^0n^{cp}(t,-\vec x))\mp \epsilon (n(t,\vec x)\pm i\gamma^0n^{cp}(t,-\vec x))=0.
\end{eqnarray}
Again we have found combinations
\begin{eqnarray}\label{26'}
\psi_{\pm}(x)=\frac{1}{2}[n(t,\vec x)\pm i\gamma^0n^{cp}(t,-\vec x)],
\end{eqnarray}
which satisfy Dirac equations with {\em different masses}, 
\begin{eqnarray}\label{27'}
[i\gamma^{\mu}\partial_{\mu}-(m\pm \epsilon)]\psi_{\pm}(x)=0.
\end{eqnarray}
We thus have
\begin{eqnarray}\label{27''}
&&n(x)=\psi_{+}(x)+\psi_{-}(x),\nonumber\\
&&n^{cp}(x)=-i\gamma^{0}[\psi_{+}(x^{0},-\vec{x})-\psi_{-}(x^{0},-\vec{x})].
\end{eqnarray}
By comparing the definition  $n^{cp}(x)=\psi^{cp}_{+}(x)+\psi^{cp}_{-}(x)$ with the second expression, we find
\begin{eqnarray}\label{28'}
&&\psi^{cp}_{\pm}(t,\vec x)=\mp i\gamma^0\psi_{\pm}(t,-\vec x),
\end{eqnarray}
which is confirmed by direct calculation using \eqref{26'} and shows the equality of particle and antiparticle masses if they are mass eigenstates~\footnote{We define C and CP by $\psi^{c}=C\bar{\psi}^{T},\ \ \bar{\psi}^{c}=\psi^{T}C$ and $\psi^{cp}=C\gamma^{0}\bar{\psi}^{T},\ \ \bar{\psi}^{cp}=\psi^{T}\gamma^{0}C$, respectively, for general fermions. Those definitions keep mass term $\bar{\psi}\psi$, for example, invariant if one uses $C^{2}=-1$ and Fermi statistics for $\psi$. One can now confirm $(\psi^{c})^{c}=CC^{T}\psi=\psi$ if one uses $C^{T}=-C$, while
$(\psi^{cp})^{cp}=C\gamma^{0}C^{T}\gamma^{0}\psi=-\psi$ and $(\bar{\psi}^{cp})^{cp}=-\bar{\psi}$. This is used to show \eqref{28'}.
}. 

Further on, we expand $\psi_\pm$ just as in the previous case in \eqref{33}.

\paragraph{General CPT-preserving case (arbitrary $\alpha$):}  When $\alpha$ is arbitrary, C and CP are simultaneously violated, with the exception of the cases discussed above. 
In this case we would follow all the steps performed in the previous situations, but solve the equations of motion in $n(x)$ and $n^{cpt}(-x)$, since the latter is always well-defined in the same Hilbert space as $n(x)$. We obtain the mass eigenfields
\begin{eqnarray}\label{32'}
\psi_{\pm}(x)=\frac{1}{2}[n(x)\pm i e^{i\alpha}i\gamma_{5}i\gamma^{2}n^{cpt}(-x)],
\end{eqnarray}
i.e.
\begin{eqnarray}\label{33'}
&&n(x)=\psi_{+}(x)+\psi_{-}(x),\nonumber\\
&&n^{cpt}(x)=-ie^{-i\alpha}i\gamma_{5}i\gamma^{2}[\psi_{+}(-x)-\psi_{-}(-x)].
\end{eqnarray}
By comparing the definition of $n^{cpt}(x)=\psi^{cpt}_{+}(x)+\psi^{cpt}_{-}(x)$ with the second expression we obtain
\begin{eqnarray}\label{34'}
\psi^{cpt}_{+}(x)=-ie^{-i\alpha}i\gamma_{5}i\gamma^{2}\psi_{+}(-x),\ \ \ \psi^{cpt}_{-}(x)=ie^{-i\alpha}i\gamma_{5}i\gamma^{2}\psi_{-}(-x).
\end{eqnarray}

\subsection*{Neutron-antineutron oscillations when P=odd}

When one discusses the neutron oscillation, the exact solution such as \eqref{33'} by itself does not help much to
understand the physical picture. First of all,  
\begin{eqnarray}\label{35}
&&[\Box +\tilde{M}^{2}]n(x)=0,\nonumber\\
&&[\Box +\tilde{M}^{2}]n^{cpt}(-x)=0,
\end{eqnarray}
do {\em not} hold for any $\tilde{M}$; consequently, the neutron and the antineutron, which are defined as the exact solutions of the CPT preserving quadratic Lagrangian \eqref{20}, cannot be on-shell for any choice of $\tilde{M}$. Related to this, if one starts with the %Majorana
field $\psi_{+}(x)$, for example, the particle stays $\psi_{+}(x)$ forever and no oscillation at all will occur. One needs to find a way to justify the superposition of two %Majorana 
particles $\psi_{\pm}(x)$ as the initial condition. In this respect, we suggest the following analogy with the {\em neutrino oscillation}: One may take the ordinary physical neutron and antineutron as an analogue of the flavor eigenstates, and the $\psi_{\pm}(x)$ particle representation as an analogue of mass eigenstates. 

To realize this picture, we assume an adiabatic switch-on of the $\Delta B=2$
interaction at $t=0$, for example. The ordinary neutron $n_{0}(x)$ described by the 1-particle wave function
\begin{eqnarray}\label{36}
\Psi_{n_{0}}(t,\vec{x})&=&\langle 0|{n}_{0}(x)a^{\dagger}(\vec{p},s)|0\rangle \nonumber\\
&=&u(\vec{p},s)e^{-ipx}
\end{eqnarray}
enters the world with ${\cal L}_{\Bslash}\neq 0$, where $a(\vec{p},s)=a(\vec{p},+,s)+a(\vec{p},-,s)$ with $m_{+}=m_{-}=m$ for the ordinary neutron. This flavor eigenstate description is then converted to a mass eigenstate description in an adiabatic manner at $t=0$,
\begin{eqnarray}\label{37}
a^{\dagger}(\vec{p},s)|0\rangle\rightarrow \big(a(\vec{p},+,s)+a(\vec{p},-,s)\big)^{\dagger}|0,\alpha\rangle,
\end{eqnarray}
with $m_{+}\neq m_{-}$, and the field is replaced by $n_{0}(x)\rightarrow n(x)$. We thus
have  
\begin{eqnarray}\label{38}
\Psi_{n}(t,\vec{x})&=&\langle 0,\alpha|{n}(x)\big(a(\vec{p},+,s)+a(\vec{p},-,s)\big)^{\dagger}|0,\alpha\rangle, \\
&=&(1/2)[u(\vec{p},+,s)e^{-i\sqrt{\vec{p}^{2}+(m+\epsilon)^{2}}t+i\vec{p}\cdot\vec{x}}+u(\vec{p},-,s)e^{-i\sqrt{\vec{p}^{2}+(m-\epsilon)^{2}}t+i\vec{p}\cdot\vec{x}}],\nonumber
\end{eqnarray} 
where $u(\vec{p},\pm,s)$ are now understood as the eigenfunctions in \eqref{31}.

After a suitable elapse of time, $t=\tau=\pi/\epsilon$ for a slow neutron, we show that the neutron is converted to an antineutron, in analogy with the flavor change in the case of neutrino oscillations. When we define the antineutron in the mass eigenstate basis, we have some freedom, namely, the C, CP or CPT conjugate of the 
neutron. The specification of the antineutron state is dictated by the symmetry of  ${\cal L}_{\Bslash}$.

For the {\bf C-conserving case}, we find a mixture 
of the antineutron component $${\cal C}\big(a(\vec{p},+,s)+a(\vec{p},-,s)\big)^{\dagger}|0,\pi/2\rangle$$  in the field operator $n(x)$,
\begin{eqnarray}\label{39}
\Psi_{n^{c}}(t,\vec{x})&=&\langle 0,\pi/2|{n}(x){\cal C}\big(a(\vec{p},+,s)+a(\vec{p},-,s)\big)^{\dagger}|0,\pi/2\rangle\nonumber\\
&=&\langle 0,\pi/2|{\cal C}^{-1}{n}(x){\cal C}\big(a(\vec{p},+,s)+a(\vec{p},-,s)\big)^{\dagger}|0,\pi/2\rangle\nonumber\\
&=&(1/2)[u(\vec{p},+,s)e^{-ipx}-u(\vec{p},-,s)e^{-ipx}].
\end{eqnarray}     
which is valid for $t\geq 0$. Here we used \eqref{29} and the invariance of the vacuum under C-conjugation.

In the {\bf CP-preserving} case, the antineutron in the mass eigenstate basis is defined  as
$${\cal CP}\big(a(-\vec{p},+,s)+a(-\vec{p},-,s)\big)^{\dagger}|0,0\rangle$$
and we find, using \eqref{27''} and $\gamma^{0}u(-\vec{p},\pm,s)=u(\vec{p},\pm,s)$,
 a mixture 
of the antineutron component in the field operator $n(x)$
\begin{eqnarray}\label{40}
\Psi_{n^{cp}}(t,\vec{x})&=&\langle 0,0|{n}(x){\cal CP}\big(a(-\vec{p},+,s)+a(-\vec{p},-,s)\big)^{\dagger}|0,0\rangle\nonumber\\
&=&\langle 0,0|({\cal CP})^{-1}{n}(x){\cal CP}\big(a(-\vec{p},+,s)+a(-\vec{p},-,s)\big)^{\dagger}|0,0\rangle\nonumber\\
&=&(i/2)[u(\vec{p},+,s)e^{-ipx}-u(\vec{p},-,s)e^{-ipx}],
\end{eqnarray}
which is valid for $t\geq 0$. 

For general $\alpha$, for which {\bf CPT is conserved}, we define the antiparticle by CPT conjugation,
 $${\cal CPT}\big(a(\vec{p},+,-s)+a(\vec{p},-,-s)\big)^{\dagger}|0,\alpha\rangle$$ and thus we have 
\begin{eqnarray}
\Psi_{n^{cpt}}(t,\vec{x})&=&\langle 0,\alpha|{n}(x){\cal CPT}\big(a(\vec{p},+,-s)+a(\vec{p},-,-s)\big)^{\dagger}|0,\alpha\rangle\\
&=&\langle 0,\alpha|({\cal CPT})^{-1}{n}(x){\cal CPT}\big(a(\vec{p},+,-s)+a(\vec{p},-,-s)\big)^{\dagger}|0,\alpha\rangle^{\star}\nonumber\\
&=&\langle 0,\alpha|-ie^{-i\alpha}i\gamma_{5}i\gamma^{2}[\psi_{+}(-x)-\psi_{-}(-x)]\big(a(\vec{p},+,-s)+a(\vec{p},-,-s)\big)^{\dagger}|0,\alpha\rangle^{\star}\nonumber\\
&=&(e^{i\alpha}/2)[u(\vec{p},+,s)e^{-ipx}-u(\vec{p},-,s)e^{-ipx}].\nonumber
\end{eqnarray}
where we used \eqref{33'} and $u(\vec{p},\pm,s)=\gamma_{5}v(\vec{p},\pm,-s)$ and $v(\vec{p},\pm,-s)=i\gamma^{2}u(\vec{p},\pm,-s)^{\star}$.

When time elapses as in \eqref{38}, we observe that in each case the antineutron component becomes the dominat one, while $\Psi_{n}(t,\vec{x})$ becomes small, due to the mass difference $m_{\pm}$, The dominance of the neutron or antineutron alternates with the passage of time, which is called {\em neutron oscillation} and occurs regardless of the C or CP properties of the $\Delta B=2$ interaction, as long as parity is violated.

\subsection{P=even,\ C=even or odd}

We next analyze the hermitian Lorentz invariant local Lagrangian consisting of the sum of \eqref{1} and \eqref{13},
\begin{eqnarray}\label{42}
{\cal L}&=&\overline{n}(x)i\gamma^{\mu}\partial_{\mu}n(x) - m\overline{n}(x)n(x)\nonumber\\
&-&\frac{i}{2}\epsilon[ e^{-i\alpha}n^{T}(x)C\gamma_{5}n(x) + e^{i\alpha}\overline{n}(x)C\gamma_{5}\overline{n}^{T}(x)],
\end{eqnarray}
but again we allow a phase for the baryon number violating parameter
\begin{eqnarray}\label{43}
\epsilon \rightarrow \epsilon e^{-i\alpha},
\end{eqnarray}
with $\epsilon>0$; $\alpha=0$ corresponds to \eqref{13} and $\alpha=\pi/2$ corresponds to the P=even, C=T=odd case in \eqref{11},
respectively. 
We may undertake a similar construction as in eq. \eqref{22}, by considering the hermitian interaction Lagrangian
\begin{eqnarray}\label{44}
{\cal L}_{int}&=&-\frac{i}{2}[n^{T}(x)C\gamma_{5}n(x)\phi(x) + \overline{n}(x)C\gamma_{5}\overline{n}^{T}(x)\phi(x)^{\dagger}],
\end{eqnarray}
where the transformation property of $\phi(x)$ is 
$\phi(x)\rightarrow e^{-2i\alpha}\phi(x)$ for $n(x)\rightarrow e^{i\alpha}n(x)$,  P=even, T=even and $\phi(x)\rightarrow \phi^{\dagger}(x)$ under C, namely, ${\cal L}_{int}$ is invariant under all the discrete symmetries. One may now consider an analogue of spontaneous symmetry breaking 
\begin{eqnarray}\label{45}
\langle 0,\alpha|\phi(x)|0,\alpha\rangle=\epsilon e^{-i\alpha},
\end{eqnarray}
to obtain the Lagrangian \eqref{42}, and the vacuum parameterized by $\alpha$ breaks the baryon number together with C and T for $\alpha\neq 0$. For $\alpha=0$, only the baryon number is broken.

We obtain the equations of motion from \eqref{42}:
\begin{eqnarray}\label{46}
&&[i\gamma^{\mu}\partial_{\mu}-m]n(x)-i\epsilon e^{i\alpha}\gamma_{5}n^{c}(x)=0,\nonumber\\
&&[i\gamma^{\mu}\partial_{\mu}-m]n^{c}(x)-i\epsilon e^{-i\alpha}\gamma_{5}n(x)=0,
\end{eqnarray}
with $n^{c}(x)\equiv C\bar{n}^{T}(x)$.

The equation \eqref{46} is solved by rewriting it as 
\begin{eqnarray}\label{47}
[i\gamma^{\mu}\partial_{\mu}-m](n(x)\pm e^{i\alpha}n^{c}(x))\mp i\epsilon \gamma_{5}(n(x)\pm e^{i\alpha}n^{c}(x))=0
\end{eqnarray}
and defining
\begin{eqnarray}\label{48}
m\pm i\epsilon \gamma_{5}=Me^{\pm 2i\theta\gamma_{5}}
\end{eqnarray}
with
\begin{eqnarray}\label{49}
M=\sqrt{m^{2}+\epsilon^{2}}.
\end{eqnarray}
Namely, we have 
\begin{eqnarray}\label{50}
[i\gamma^{\mu}\partial_{\mu}-M]e^{\pm i\theta\gamma_{5}}(n(x)\mp e^{i\alpha}i\gamma_{5}i\gamma^{2}n^{cpt}(-x))=0
\end{eqnarray}
where we replaced $n^{c}(x)\rightarrow -i\gamma_{5}i\gamma^{2}n^{cpt}(-x)$ since $n^{c}(x)$ is not defined in the Hilbert space of $n(x)$ for $\alpha\neq 0$.  
We thus identify the combinations
\begin{eqnarray}\label{51}
\psi_{+}&=&\frac{1}{2}e^{i\theta\gamma_{5}}(n(x)- e^{i\alpha}i\gamma_{5}i\gamma^{2}n^{cpt}(-x)),\nonumber\\
\psi_{-}&=&\frac{1}{2}e^{- i\theta\gamma_{5}}(n(x)+ e^{i\alpha}i\gamma_{5}i\gamma^{2}n^{cpt}(-x)),
\end{eqnarray}
which satisfy the standard Dirac equation
\begin{eqnarray}\label{52}
[i\gamma^{\mu}\partial_{\mu}-M]\psi_{\pm}=0.
\end{eqnarray}
One can confirm that 
\begin{eqnarray}\label{53}
&&\psi_{\pm}^{p}(x^{0},\vec{x})=\gamma^{0}\psi_{\mp}(x^{0},-\vec{x}).
\end{eqnarray}
%We also have 
%\begin{eqnarray}\label{54}
%(e^{\mp i\theta\gamma_{5}}\psi_{\pm})^{c}(x)= e^{\mp i\theta\gamma_{5}}\psi_{\pm}(x)
%\end{eqnarray}
Thus we have the exact solutions of the field equations \eqref{46},
\begin{eqnarray}\label{55}
&&n(x)=[e^{-i\theta\gamma_{5}}\psi_{+}(x)+e^{i\theta\gamma_{5}}\psi_{-}(x)],\nonumber\\
&&n^{cpt}(x)=-e^{-i\alpha}i\gamma_{5}i\gamma^{2}[e^{-i\theta\gamma_{5}}\psi_{+}(-x)-e^{i\theta\gamma_{5}}\psi_{-}(-x)].
\end{eqnarray}
When one defines $\psi_{N^\pm}(x)$ with a shifted mass $M=\sqrt{m^{2}+\epsilon^{2}}$ by
\begin{eqnarray}\label{56}
&&\psi_{N^{\pm}}(x)\equiv \psi_{+}(x)\pm \psi_{-}(x),\nonumber\\
&&\psi^{p}_{N^{\pm}}(x)=\pm \gamma^{0}\psi_{N^{\pm}}(x^{0},-\vec{x}),
\end{eqnarray}
 one can rewrite \eqref{55} as
\begin{eqnarray}\label{57}
&&n(x)=[\cos\theta \psi_{N^{+}}(x)-\sin\theta (i\gamma_{5})\psi_{N^{-}}(x)], \nonumber\\
&&n^{cpt}(x)=-e^{-i\alpha}i\gamma_{5}i\gamma^{2}[\cos\theta \psi_{N^{-}}(-x)-\sin\theta (i\gamma_{5})\psi_{N^{+}}(-x)].
\end{eqnarray}
The "neutron" $n(x)$, which  is written as a superposition of $\psi_{N^{+}}$ and $i\gamma_{5}\psi_{N^{-}}(x)$, has a well-defined mass $M$. To the order linear in $\epsilon$, we have $M=m$ and 
\begin{eqnarray}\label{66'}
n(x)\simeq [n_{0}(x)-\theta e^{-i\alpha}(i\gamma^{2})n_{0}^{cpt}(-x)],
\end{eqnarray}
namely, the new field $n(x)$ is a superposition of the original neutron $n_{0}(x)$ and antineutron $n_{0}^{cpt}(-x)$.

In the present case, we have no oscillation because of the degeneracy of the masses of the fields $\psi_{+}(x)$ and $\psi_{-}(x)$, but the expression in \eqref{66'} shows that one observes both the decay $n\rightarrow p+e^{-}+\bar{\nu}_{e}$ and the decay $n\rightarrow \bar{p}+e^{+}+\nu_{e}$ through a small mixture of $(i\gamma^{2})n_{0}^{cpt}(-x)$.  Also, the pair annihilation of the neutron takes place when it collides with a bulk matter. 

The difference in physical implications of the presence of oscillation and its absence is that, if the oscillation should take place, the decay $n\rightarrow \bar{p}+e^{+}+\nu_{e}$, for example,  would happen exclusively if one observes the neutron at the proper moment of complete oscillation, while we do not have any such  "bunching effect" without the oscillation.

\section{Discussion}
The analysis in~\cite{berezhiani} is very stimulating, but its conclusion that the neutron
oscillation {\em per se} inevitably implies CP violation is shown in our analysis not to be warranted.  

One can confirm that, if one adds the P=odd and P=even ${\cal L}_{\Bslash}$ terms both with $\alpha=0$ in \eqref{20} and \eqref{42} and if one replaces $n(x)\rightarrow \nu(x)$, one obtains the {\em CP preserving} right-handed neutrino mass term,
\begin{eqnarray}
{\cal L}_{\nu - mass}=
-\frac{i}{2}m_{R}[\nu^{T}(x)C(1+\gamma_{5})\nu(x) - \overline{\nu}(x)(1-\gamma_{5})C\overline{\nu}^{T}(x)].
\end{eqnarray}
This fact implies that the notion of charge conjugation is ill-defined in the presence of the baryon number 
non-conservation and that the  CP violation in the neutron oscillation is generally spurious and may not be used in the physical analysis of baryogenesis, just as the right-handed neutrino mass term does not provide CP violation for leptogenesis. Note that the CP property of the above neutrino mass term can be freely changed if one performs a neutrino number phase transformation.
However, there is a qualification to the above statement; one cannot completely eliminate the possible CP violation if both P=odd and P=even $\Delta B=2$ terms with arbitrary phases exist in the Lagrangian, although neutron oscillation {\em per se} takes place only with P=odd terms. This complication in CP properties is related to the fact that CP properties of the left-right symmetric theory, on which the original suggestion of the neutron oscillation is based~\cite{mohapatra}, are different from those of the Standard Model. 

In contrast, CPT is always intact in the local Lorentz invariant  hermitian Lagrangian, but CPT symmetry does not necessarily imply that the neutron and the antineutron defined as the exact solutions of the quadratic Lagrangian satisfy the free Dirac equation with a well-defined mass in the presence of a $\Delta B=2$ interaction.

We have also shown that the neutron oscillation in a proper sense does not take place if parity is conserved in a Lorentz invariant local effective Lagrangian even with a $\Delta B=2$ term. Phenomenologically, this implies that the signals characteristic to {\em oscillation} of the neutron are not observed in parity conserving theory, although the neutron number violating transition itself can take place. 

We discussed the issue of neutron oscillation by assuming CPT invariance; the possible CPT violation in the hadron sector appears to be very small as is indicated by both experimental limit $|m_{K}-m_{\bar{K}}|<0.44\times 10^{-18}$ GeV~\cite{particledata} 
 and a recent model study within an extension of the Standard Model~\cite{fujikawa-tureanu}.

\subsection*{Acknowledgments}

We thank Masud Chaichian for very helpful discussions. This work is supported in part by JSPS KAKENHI (Grant No. 25400415) and the Vilho,Yrj\"o and Kalle V\"ais\"al\"a Foundation. The support of the Academy of Finland under the
Projects no. 136539 and 272919 is gratefully acknowledged.

\appendix 
\section{Notational convention}
We here summarize the definitions of various discrete transformation rules
for a Dirac fermion $\psi(x)$.
We follow the Bjorken--Drell convention~\cite{bjorken} with the metric $g^{\mu\nu}=(1,-1,-1,-1)$ and $\{\gamma^{\mu},\gamma^{\nu}\}=2g^{\mu\nu}$, but our choice of spinor solutions, which includes a factor $\sqrt{m/E}$, is given by
\begin{eqnarray}\label{60}
&&u(\vec{p},s)=\sqrt{\frac{E+m}{2E}}\left(\begin{array}{c}
            \xi(s)\\
            \frac{\vec{\sigma}\cdot\vec{p}}{E+m}\xi(s)
            \end{array}\right), \ \ \
v(\vec{p},s)=\sqrt{\frac{E+m}{2E}}\left(\begin{array}{c}
            \frac{\vec{\sigma}\cdot\vec{p}}{E+m}\xi(-s)\\
            \xi(-s)
            \end{array}\right)
\end{eqnarray}
with a two-component spinor $\xi(\pm 1)$ defined at the rest frame.
\\
The parity transformation is defined by
\begin{eqnarray}\label{61}
   &&\psi(t,\vec{x})\to\gamma_0 \psi(t, -\vec{x}),\quad
   \overline \psi(t,\vec{x})\to\overline \psi(t, -\vec{x})\gamma_0.
\end{eqnarray} 
The charge conjugation is defined by
\begin{eqnarray}\label{62}
   &&\psi(x)\to-C^{-1}\overline \psi^T(x)=C\overline \psi^T(x),\quad
   \overline \psi(x)\to \psi^T(x)C,
\end{eqnarray}
with $C=i\gamma^{2}\gamma^{0}$
which satisfies 
\begin{equation}\label{63}
   C^\dagger C=1,\quad C^T=-C,\quad C\gamma_\mu C^{-1}=-\gamma_\mu^T,
   \quad C\gamma_5C^{-1}=\gamma_5^T.
\end{equation}
The (anti-unitary) time reversal is defined by 
\begin{eqnarray}\label{64}
{\cal T}\psi_{\alpha}(t,\vec{x}){\cal T}^{-1}=T_{\alpha\beta}\psi_{\beta}(-t,\vec{x}),\quad
{\cal T}\psi^{\dagger}_{\alpha}(t,\vec{x}){\cal T}^{-1}=\psi^{\dagger}_{\beta}(-t,\vec{x})(T^{-1})_{\beta\alpha},
\end{eqnarray}
with \ $T=i\gamma^{1}\gamma^{3}, \ \  T\gamma_{\mu}T^{-1}=\gamma_{\mu}^{T}=(\gamma^{\mu})^{\star},\ \
T=T^{\dagger}=T^{-1}=-T^{\star}$.\\
The (anti-unitary) CPT is defined by 
\begin{eqnarray}\label{65}
&&{\cal CPT}\psi_{\alpha}(t,\vec{x})({\cal CPT})^{-1}=i\gamma^{5}_{\alpha\beta}\psi^{\dagger}_{\beta}(-t,-\vec{x}),\nonumber\\
&&{\cal CPT}\bar{\psi}_{\alpha}(t,\vec{x})({\cal CPT})^{-1}=-i\psi_{\beta}(-t,-\vec{x})(\gamma^{5}\gamma_{0})_{\beta\alpha}.
\end{eqnarray}

\end{document}